\documentclass[letterpaper,10pt,conference]{IEEEtran}
\usepackage[latin1]{inputenc}
\usepackage{epsf}\epsfverbosetrue
\usepackage{graphics,epsfig,color}
\usepackage{graphicx,epsfig}
\usepackage{multirow}
\usepackage{slashbox}
\usepackage{alltt}
\usepackage{subfigure}
\usepackage{color}
\usepackage{float}
\usepackage{url}
\usepackage{graphicx}
\usepackage{verbatim} 
\usepackage{footnote} 
\usepackage{latexsym} 
\usepackage{amsmath}
\usepackage{sidecap} 
\usepackage{color}
\usepackage{wrapfig}
\usepackage{algorithm}
\usepackage{eso-pic}
\usepackage{fix-cm}

\AddToShipoutPicture*{\small \sffamily\raisebox{1.8cm}{\hspace{1.8cm}978-1-4244-7265-9/10/\$26.00 \copyright2010 IEEE}}

\begin{document}
%
\title{\fontsize{20}{22}\selectfont Channel Assortment Strategy for Reliable Communication in Multi-Hop Cognitive Radio Networks}

%
%
%
%
%

%

\author{Mubashir Husain Rehmani \\
\normalsize {Lip6/Universite Pierre et Marie Curie - Paris 6, France. E-mail: mubashir.rehmani@lip6.fr}  \\
}

\maketitle




\vspace{-0.4cm}
\section{Introduction}
\label{sec:in}
Channel selection plays a vital role in {\it efficient and reliable data dissemination}. In the
context of Cognitive Radio Network (CRN)~\cite{survey}, channel selection is more challenging due to the traffic pattern and channels' occupancy of primary radio (PR) nodes. Moreover, Cognitive Radio (CR) transmissions should not degrade the reception quality of PR nodes and should be immediately interrupted
whenever a neighboring PR activity is detected~\cite{hicham}.
Thus, it is essential and, however, extremely challenging for CR nodes to correctly select channels allowing reliable communication. 

In contrast with previous approaches which focused on single-hop wireless
infrastructure, the main problem we tackle here is {\it the channel assortment to undergo efficient and reliable data dissemination throughout a
multi-hop
CRN}. Basically, we want to increase the transmission coverage of CR nodes, and consequently, the global network dissemination coverage, by selecting good
qualified channels for communicating. More specifically, our channel assortment strategy, named `SURF' (an allusion for riding on channels),
empowers CR nodes with the ability to infer, based on information regarding PR occupancy, the less
PR-occupied channel to use. 
On the flip side, although less occupied channel by PR nodes provides higher
space sharing to CR nodes, if all CR nodes decide to switch to it, a higher CR competition
will be perceived, increasing the probability of collision and message losses. In this way, {\it how to find
a good compromise between transmission opportunity in terms of PR occupancy and the number of CR neighbors, constitutes
the main goal of this paper}.

\vspace{-0.2cm}
\section{Channel Assortment strategy}
\label{sec:proposal}
\vspace{-0.1cm}We consider a multi-hop CRN that has a set of: PR nodes, CR nodes, and
network links connecting CR nodes. We assume that a set of frequency channels is available in the network and
may be used by PR or CR nodes. We consider a slotted-based channel in which
each channel is divided into total number of $\tau_t=\tau_o + \tau_a$ time slots, where $\tau_o$ and $\tau_a$ are the slots
occupied by PR nodes and the available slots, respectively.
In essence, in order to be able to communicate in a CRN, CR nodes must be
cognitively aware of their surroundings and able to make decisions about ``when, where, and how'' to
communicate. The SURF goal is to support nodes in their decision about ``where'' to transmit or overhear,
i.e. which channel to use. Here, the role of overhearing is for receiving data.

In a multi-hop context, at each hop, CR nodes will select for {\it data dissemination} the
best channel in the channel classification provided by SURF. 
In order to limit energy consumption and
communication overhead, the SURF assortment is performed in {\it a distributed way} and is based only on
{\it information locally inferred by CR nodes}. In fact, by implementing the same strategy at the sender and the receiver, SURF helps both of them tune to the appropriate channel for undergoing transmissions or reception without the need of any prior information exchange or synchronization. 


Consider the set $C$ of total frequency channels. Since, the number of channels changes with time and location due to PR activity, each CR node has its own available channel set among total frequency channels $C$, which we consider provided by the spectrum sensing block~\cite{arslan}. We call this available channel set as $Acs$, where $\forall{\it Acs} \in{ {\it C}}$. $Acs$ describes the channels set a CR node has access to. After sensing them and according to the SURF strategy,
each CR node classifies the sensed channels according to their availabilities and selects for transmission
and/or overhearing the best weighted channel. SURF assigns weights for channels according to the
Eq.~\ref{pequation}: \vspace{-0.2cm}
\begin{equation}
\label{pequation} {\forall {\it i} \in {\it C}: P_w^{(i)} = e^{-PR_o^{(i)}} \times CR_o^{(i)} }
\end{equation}

\vspace{-0.25cm}
$P_w^{(i)}$ describes the availability level of a channel $(i)$ and is

\noindent calculated based on the occupancy of PR (i.e. $PR_o^{(i)}$) and CR (i.e. $CR_o^{(i)}$) nodes in this channel. If a CR node finds
two or more channels having identical values of $P_w$ then it selects the one that has lower $PR_o^{(i)}$. If the problem persists,
then, a channel is randomly selected among them.

\begin{figure*}[htbp]
  \begin{center}
    \subfigure[]
    {
      \label{fig0}
      \epsfxsize= 6.7cm
	  \leavevmode\epsfbox{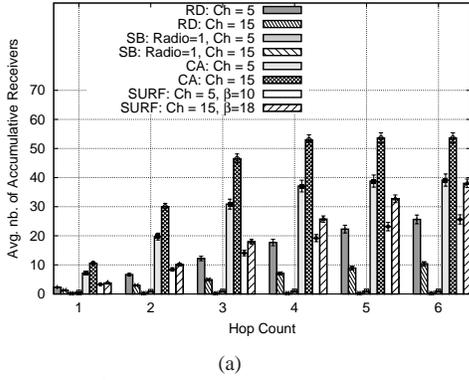}
    }\hspace{1cm}
  \subfigure[]
    {
      \label{fig1}
      \epsfxsize= 6.7cm
      \leavevmode\epsfbox{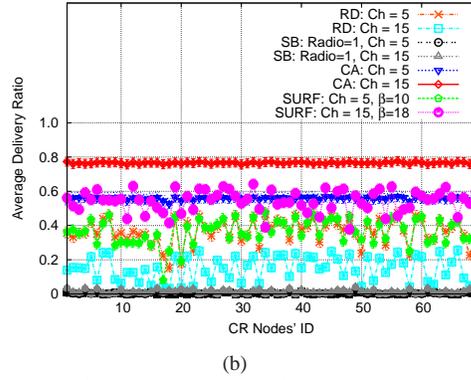}
    }\hspace{1cm}
   \vspace{-0.3cm} \caption{Comparison of random (RD), selective broadcasting (SB), centralized authority (CA) and our strategy (SURF), when {\it channels=5} and {\it channels=15} in a CRN with 70 CR nodes. (a) Hop Count and average number of accumulative receivers. (b) CR Nodes' ID and Average delivery ratio.}\vspace{-0.8cm}
    \label{fig:config}
  \end{center}
\end{figure*}

We rely on methods in~\cite{arslan} and assume that the spectrum opportunity map of PR nodes and CR neighbor ($CR_n^{(i)}$) information is
available to CR nodes. Note that $CR_n^{(i)}$ is the number of potential CR neighbors competing for the same channel.
This map is then used to compute the percentage of the channel $i$ occupied by PR nodes (i.e., the ratio between the number of PR nodes and the total number of time slots $\tau_t$), or
$PR_o^{(i)}$. 
The remaining available percentage of the channel $i$, i.e. $1 - PR_o^{(i)}$, gives then the
spectrum opportunity available for channel sharing among CR nodes, named $CR_{as}^{(i)}$. 

If a central entity could control the channel access for CR nodes, using a channel with higher available
space could allow achieving higher number of 1-hop CR receivers.
Since no central entity is
present, there is no way to prevent collision and message losses when the number of CR nodes competing for
the same channel increases. To cope with this and to allow nodes selecting channel having a good compromise
between the number of CR neighbor receivers and the number of competing CR transmitters, we use a {\it
Tenancy Factor}, named $\beta$. $\beta$ is used to compute the CR occupancy $CR_o^{(i)}$ of each channel $i$. $\beta$
provides the upper bound in terms of number of CR neighbors on a particular channel for which the communication
is still performed with a good probability of success. The goal is then, to maximize the chances of
selecting channels that have higher but enough number of CR neighbors (close to $\beta$) allowing reliable communication.


Algorithm~\ref{Alg} shows how
CR nodes calculates CR occupancy according to $\beta$ and the number of CR neighbors
competing for the channel.
We performed extensive simulations in order to determine the appropriate value of
${\it \beta}$ to be used. We ask readers to refer to~\cite{tr} for more details.

 \begin{algorithm}[t]
  \caption{\footnotesize{Algorithm for CR occupancy's computation}}\label{Alg}
\scriptsize
\textbf{if} $CR_n^{(i)} \textless  \beta$ \\
     \textbf{then}  $ CR_o^{(i)} \leftarrow \tau_a / (\beta - CR_n^{(i)})$\\
\textbf{else if} $CR_n^{(i)} = \beta$ \\
    \textbf{then} $ CR_o^{(i)} \leftarrow CR_{as}^{(i)} $ \\
\textbf{else} $CR_n^{(i)} \textgreater \beta$ \\
    \textbf{then}  $ CR_o^{(i)} \leftarrow \tau_a / \tau_t . CR_n^{(i)}$ \\
\textbf{end if}
\end{algorithm}

\vspace{-0.20cm}
\section{Performance Analysis}
\label{sec:analysis}
%
\vspace{-0.05cm}Results are generated from an average of 1000 simulations, 
with 95\% of confidence intervals. We consider 30 PR nodes; uniformly
distributed among the existing channels and $\tau_t$=6 total time slots for each channel. 
The transmission range is set to {\it R} = 250m and
the average CR neighbor density $d_{avg}$ set to 20. The
number of CR nodes is fixed to {\it N=70} and randomly deployed within a square area of {\it $a^2=$}707x707$m^{2}$. 
{\it TTL} is used as the maximum number of hops
required for a packet to traverse the whole network and is set to $\lceil \dfrac{2a}{R} \rceil$, i.e. $TTL=6$.
The $Acs$ size is then set to 3 and 8 for 5 and 15 existing channel ($Ch$), respectively. The appropriate value of $\beta$ is set to 10 and 18 when using $Ch=5$ and $Ch=15$ respectively, which provides the best tradeoff
among number of receivers and loss ratio (cf.~\cite{tr}). 

We compare SURF with a random strategy (RD) and two variants of selective broadcasting strategy~\cite{agrawal}:
without (SB) and with (CA) centralized authority.
CA can be thus used as an upper bound in message
dissemination comparison, as it maximizes the number of receptions by performing overhearing over multiple channels,
simultaneously. 
Since, our goal is to efficiently disseminate the data, therefore, we have chosen two performance metrics to compare these four strategies: (i) the average delivery ratio, which is the ratio of packet received by a
particular CR node over total packets sent in the network and (ii) the average number of accumulative CR receivers per hop.

Fig.~\ref{fig0} compares the number of accumulative CR receivers at each hop of communication (i.e until {\it TTL=0}) for the four strategies. SURF allows the message dissemination to 55\% of nodes in the network. It can be clearly seen that SURF outperforms RD and SB and compared to CA, only provides a decrease of 25\% in performance (1.5 less transmissions). Note that CA have higher number of receivers at the cost of multiple transmissions (2.5 transmissions), while in RD and SURF, each CR node transmits only once.

Fig.~\ref{fig1} compares delivery ratio of RD, SB, CA, and SURF, as a function of CRs' ID. SURF increases considerably the
delivery ratio compared to RD and SB and only reduces in 20\% the performance compared to CA. In particular, for {\it Ch=5} and {\it Ch=15}, SURF guarantees the delivery of approximately 60\% of messages (single transmission), contrarily to less
than 20\% for the RD and SB (single and 2.5 transmissions, resp.) and 80\% for CA (2.5 transmissions). These results show the good level of network connectivity provided by SURF, suitable for reliable dissemination.


\vspace{-0.25cm}

\section{Future Works}
\label{sec:conclusion}
\vspace{-0.18cm}
We intend in future
to investigate SURF performance under dynamic traffic by consideration of data rates and traffic volume generated by CR nodes and optimize data dissemination delay. Moreover,
prediction and history can also be accounted in SURF to enhance the performance.\vspace{-0.17cm}

\section{Acknowledgements}
\vspace{-0.12cm}The author would like to thank Aline C. Viana, Hicham Khalife, and Serge Fdida for the helpful discussions and suggestions, and for their
very careful reading of the manuscript.\vspace{-0.20cm}


\bibliographystyle{abbrv}

\begin{thebibliography}{1}
{\scriptsize \vspace{-0.24cm}
\bibitem{survey}
I.~F. Akyildiz, W.-Y. Lee, M.~C. Vuran, and S.~Mohanty, ``Next
  generation/dynamic spectrum access/cognitive radio wireless networks: a
  survey,'' \emph{Computer Networks}, vol. 50 , Issue 13, pp. 2127 -- 2159, 2006.



\bibitem{hicham}
H.~Khalife, S.~Ahuja, N.~Malouch, and M.~Krunz, ``Probabilistic path selection
  in opportunistic cognitive radio networks,'' in \emph{IEEE
  globecom conference}, 2008.

%




\bibitem{arslan}
T.~Yucek and H.~Arslan, ``A survey of spectrum sensing algorithms for cognitive
  radio applications,'' \emph{IEEE Communications Surveys and Tutorials}, vol.
  11, Issue No. 1, pp. 116--130, First Quarter 2009.



\bibitem{agrawal}
Y.~R. Kondareddy and P.~Agrawal, ``Selective broadcasting in multi-hop
  cognitive radio networks,'' in \emph{IEEE Sarnoff Symposium},
  2008.

\bibitem{tr}
M. H. Rehmani, A. C. Viana, H. Khalife,	and S. Fdida, ``Toward Reliable Contention-aware Data Dissemination in Multi-hop Cognitive Radio Ad Hoc Networks'', INRIA RR-0375, 2009, France. http://hal.inria.fr/inria-00441892/en/.









}
\end{thebibliography}


\end{document}